\documentstyle[preprint,aps,epsfig]{revtex}
\newcommand{\lsim}{\mathrel{\mathop{\kern 0pt \rlap
  {\raise.2ex\hbox{$<$}}}
  \lower.9ex\hbox{\kern-.190em $\sim$}}}
\newcommand{\gsim}{\mathrel{\mathop{\kern 0pt \rlap
  {\raise.2ex\hbox{$>$}}}
  \lower.9ex\hbox{\kern-.190em $\sim$}}}

\newcommand{\beq}{\begin{equation}}
\newcommand{\eeq}{\end{equation}}

\newcommand{\W}{{\mathcal W}}

\begin{document}
\draft
%\preprint{\vbox{\hbox{AS-ITP-2004-}
      %\hbox{hep-ph/0403}
%}}
\title{$[SU(3)\times SU(2)\times U(1)]^2$ and Strong Unification}

\author{Chun Liu}

\vspace{1.5cm}

\address{Institute of Theoretical Physics, Chinese Academy of Sciences,\\
P. O. Box 2735, Beijing 100080, China\\}

\maketitle
\thispagestyle{empty}
\setcounter{page}{1}
\begin{abstract}
  A supersymmetric model with gauge symmetry $G_1\times G_2$, where 
$G_i=SU(3)_i\times SU(2)_i\times U(1)_i$, is constructed within the 
framework of gauge mediated supersymmetry breaking.  At the energy 
scale $\sim (10-100)$ TeV where the gauge symmetry breaks down to the 
Standard Model (SM), $G_1$ is strong and $G_2$ is weak.  The observed 
gauge coupling constant unification of the SM is attributed to that of 
$G_2$.  The messenger fields and Higgs fields just satisfy the 
condition that makes $G_2$ a realization of strong unification.  The 
SM gauginos are predicted to be generally heavier than the sleptons 
and squarks.  

\end{abstract}
\vspace{1.5cm}

\hspace{1.1cm}Keywords: gauge interaction, supersymmetry.


\newpage

Extensions of the SM aim at understanding new experimental results or 
unsolved theoretical problems.  The most popular approach is the grand 
unification theories (GUTs) \cite{gut}, such as the $SU(5)$ GUT.  There 
are indirect experimental evidences for GUTs from LEP and neutrino 
physics.  To make GUTs viable, supersymmetry  (SUSY) \cite{susy,susy1} 
is a must.  One of the novel idea towards GUT is the so-called strong 
unification \cite{strong1,strong2}.  In the strong GUT, the SM gauge 
coupling constants just reach their common Landau pole at the 
unification energy scale.  

Strong GUT is interesting not only due to its novelty, but also because 
of its usefulness.  There is a discrepancy between the measured value 
of the QCD strong coupling constant at $M_Z$, which is 
$\alpha_s^{\rm exp}(M_Z)\simeq 0.1172\pm 0.002$ \cite{pdg}, and that 
predicted by the minimal SUSY SM (MSSM) 
$\alpha_s^{\rm MSSM}(M_Z)\simeq 0.126$.  The discrepancy is reduced if 
extra matters are added into the MSSM.  To keep the unification, the 
additional states should be in complete representation of GUT gauge 
groups.  To the two-loop level, it has been shown \cite{strong2}, for 
example, that $\alpha_s(M_Z)\simeq 0.1163$ if there are additional six 
multiplets in ${\bf 5}+{\bf \bar{5}}$ under $SU(5)$ with masses of 
$\sim 214$ TeV.  However, the model would be artificial if these 
additional matters are naively added.  

We will illustrate that a SUSY model with the gauge symmetry 
$SU(3)_1\times SU(2)_1\times U(1)_1\times SU(3)_2\times SU(2)_2\times U(1)_2$ 
can be a nontrivial realization of the strong GUT. There are multiple 
motivations to consider such an extension of the SM 
\cite{other,ckss,kl}.  In Ref. \cite{kl}, such kind of models were 
proposed as GUT generalization of the SUSY top-color 
\cite{susytopcolor}.  They provide a solution to the SUSY flavor 
changing neutral current problem.  However, the gauge coupling constant 
behavior was rather bad at high energies because of the introduction 
of too many extra matter fields which made the gauge interactions to be 
too much strong.  This situation brings us to further think of their 
connection with the idea of the strong GUT.  In this paper, after 
naturally modifying the Higgs and messenger contents of the model, we 
note that the extra matters additional to the MSSM can make the SM-like 
gauge interaction $SU(3)_2\times SU(2)_2\times U(1)_2$ a strong GUT.  

We consider a SUSY theory with the gauge group $G_1\times G_2$ in the 
framework of gauge mediated SUSY breaking (GMSB) \cite{gmsb}, where 
$G_i=SU(3)_i\times SU(2)_i\times U(1)_i$ ($i=1,2$).  The three 
coupling constants of $G_1$ are large, and those of $G_2$ are small at 
the TeV scale.  The three generations of matter carry nontrivial 
quantum numbers of $G_2$ only.  These numbers are assigned in the same 
way as they are under the SM gauge group.  One gauge singlet chiral 
superfield $X$ is introduced for SUSY breaking, 
\beq
\label{1}
\langle X_s\rangle\neq 0\,,~~~\langle F_X\rangle\neq 0\,,
\eeq
with $X_s$ and $F_X$ being the scalar and auxiliary components of $X$.  
The vacuum expectation values are taken to be real.  

For the SUSY breaking messengers and gauge symmetry breaking Higgs', 
it is easy to consider them through imaging global $SU(5)_i$ symmetry 
into which $G_i$ is embedded.  The messengers with their quantum 
numbers under $SU(5)_1\times SU(5)_2$ are 
\beq
\label{2}
\begin{array}{ll}
T_1(5, 1)\,,&~~~\bar{T_1}(\bar{5}, 1)\,,\\
T_2(1, 5)\,,&~~~\bar{T_2}(1, \bar{5})\, 
\end{array}
\eeq
The relevant superpotential is 
\beq
\label{3}
\W_1=c_1XT_1\bar{T_1} + c_2XT_2\bar{T_2}\,,
\eeq
where $c_1$ and $c_2$ are coupling constants of order one.   The fields 
$T_i$ and $\bar{T_i}$ are massive at tree level.  Their fermionic 
components compose a Dirac fermion with mass $c_i\langle X_s\rangle$, 
while the scalar components have a squared-mass matrix 
\beq
\label{4}
(T_{is}^*~~\bar{T_{is}})
\left(
\begin{array}{cc}
c_i^2\langle X_s\rangle^2 & c_i\langle F_X\rangle    \\
c_i\langle F_X\rangle    & c_i^2\langle X_s\rangle^2 \\
\end{array}
\right)
\left(
\begin{array}{c}
T_{is} \\
\bar{T_{is}}^*
\end{array}
\right)\,.
\eeq
The mass eigenstates and squared-mass eigenvalues are 
\beq
\label{5}
\begin{array}{c}
\displaystyle\frac{1}{\sqrt{2}}(T_{is}+\bar{T_{is}}^*)~~~ {\rm with}~~~ 
m_{i1}^2=c_i^2\langle X_s\rangle^2+c_i\langle F_X\rangle\,,\\[3mm]
\displaystyle\frac{1}{\sqrt{2}}(T_{is}-\bar{T_{is}}^*)~~~ {\rm with}~~~ 
m_{i2}^2=c_i^2\langle X_s\rangle^2-c_i\langle F_X\rangle\,.
\end{array}
\eeq
It is assumed that $c_i\langle F_X\rangle <c_i^2\langle X_s\rangle^2$.  
Because $\langle F_X\rangle\neq 0$, SUSY breaking occurs in the fields 
$T_i$'s and $\bar{T_i}$'s at tree-level.  $G_1$ and $G_2$ sectors get 
to be soft SUSY breaking via the messengers at loop level.  Because 
$G_2$ is weak at TeV scale, its SUSY breaking effects can be calculated 
perturbatively, for example, $G_2$ gaugino soft masses are 
\beq
\label{6}
M_{\lambda'_r} \simeq \frac{\alpha'_r}{4\pi}
\frac{\langle F_X\rangle}{\langle X_s\rangle}\,, 
\eeq  
where $\alpha'_r=g'_r/4\pi$ with $g'_r$ being the gauge coupling 
constants of $G_2$.  And $r=1,2,3$ corresponding to the groups $U(1)$, 
$SU(2)$, and $SU(3)$, respectively.  However, $G_1$ is strong, we can 
only estimate its gaugino masses 
\beq
\label{7}
M_{\lambda_r}\simeq\frac{\langle F_X\rangle}{\langle X_s\rangle}\,.
\eeq
Numerically the messenger masses are about  $(10-100)$ TeV.  

A pair of Higgs $\Phi_1(5,\bar{5})$ and $\Phi_2(\bar{5},5)$ breaks the 
$G_1\times G_2$ gauge symmetry down to that of the SM.  One gauge 
singlet superfield $Y$ is introduced for the gauge symmetry breaking.  
The superpotential of them is written as follows,
\beq
\label{8}
\W_2=c'Y[{\rm Tr}\,(\Phi_1\Phi_2)-\mu'^2]\,,
\eeq
where the trace is taken with regard to both $SU(3)_1\times SU(3)_2$ 
and $SU(2)_1\times SU(2)_2$.  $\mu'$ is the energy scale relevant to 
the gauge symmetry breaking, and $c'$ is the coupling constant.  The 
Higgs fields get soft masses like that given in Eq. (\ref{7}).  However, 
the above superpotential is not enough to guarantee all the $\Phi_i$ 
fermion components to be massive.  Their masses are nonvanishing when 
a superfield $A$ which is in adjoint representation of $SU(5)_1$ is 
introduced with the following superpotential \cite{ckss}, 
\beq
\label{9}
\W'_2=c'_2{\rm Tr}\,(\Phi_2 A \Phi_1)\, 
\eeq
with $c'_2$ being the coupling constant.  The details of the gauge 
symmetry breaking go the same way as that in Ref. \cite{kl} (its Eqs. 
(10-17)).  The VEVs of the $\Phi_i$'s are given as 
\beq
\label{10}
\langle\Phi_{1_s}\rangle=\langle\Phi_{2_s}\rangle=v I_3\otimes I_2\,, 
\eeq
where $I_3$ and $I_2$ are the unit matrices in the space of 
$SU(3)_1\times SU(3)_2$ and $SU(2)_1\times SU(2)_2$, respectively.  The 
coupling constants of the SM $SU(3)_c\times SU(2)_L\times U(1)_Y$ are 
\beq
\label{11}
\displaystyle \frac{1}{g_s^2} = \frac{1}{g_3^2}+\frac{1}{g_3'^2}\,,~~~ 
\frac{1}{g^2}   = \frac{1}{g_2^2}+\frac{1}{g_2'^2}\,,~~~ 
\frac{1}{g'^2}  = \frac{1}{g_1^2}+\frac{1}{g_1'^2}\,.
\eeq
Numerically, the gauge symmetry breaking scale $v$ is about $(10-100)$ 
TeV.  

Electroweak symmetry breaking is achieved via a pair of Higgs 
superfields $H_u$ and $H_d$ which are nontrivial only under $G_2$ 
\cite{kl}.  

Around $10-100$ TeV, there are many matter fields which will run the 
gauge coupling constants to be large at high energies.  The matter 
fields introduced additional to MSSM are complete $SU(5)$ multiplets.  
Therefore, the unification scale $3\times 10^{16}$ GeV is still the 
same as that of the MSSM, but the values of the coupling constants are 
significantly different.  This model is a candidate of strong GUT.  

Below the scale $v$, the $G_1\times G_2$ breaks spontaneously down to 
the MSSM.  From Eq. (\ref{11}), it is easy to see that the gauge 
coupling constants of MSSM are almost fully determined by that of $G_2$, 
because $g_i\gg g'_i$.  Therefore the observed unification of MSSM is 
attributed to the unification of $G_2$.  

Above $(10-100)$ TeV scale, the theory is $G_1\times G_2$.  As far as 
the $G_2$ sector is concerned, the new matter fields in addition to the 
MSSM are the messengers $T_1$ and $\bar{T_1}$, and Higgs fields 
$\Phi_1$ and $\Phi_2$.  The messenger fields compose one 
${\bf 5}+{\bf \bar{5}}$ multiplet with a mass $c\langle X_s\rangle$ and 
the Higgs' contribute five ${\bf 5}+{\bf \bar{5}}$ multiplets with 
masses $c'v$ as well as $c'_2v$.  We have the freedom to adjust all the 
masses of these six ${\bf 5}+{\bf \bar{5}}$ multiplets to be about 
$214$ TeV.  As has been shown in Ref. \cite{strong2}, the gauge 
couplings reach their common Landau pole at the GUT scale 
$\sim 3\times 10^{16}$ GeV.  Namely in this case, $G_2$ is a 
realization of the strong GUT.  

Some remarks are necessary.  (1) The perturbative calculation in Ref. 
\cite{strong2} was not reliable around the GUT scale because of the 
large coupling constants.  But around $100$ TeV where the perturbative 
domain lies, its reliability was under control.  It is in the latter 
low energy region where we have made use of Ref. \cite{strong2}.  (2) 
On the other hand, $G_1$ sector is also expected to be a GUT.  
$(10-100)$ TeV is already its non-perturbative region, we have no 
reliable method yet to make detailed analysis.  (3) The unification 
simply means that the gauge coupling constants are equal at certain 
scales.  We have not introduced any unified gauge group.  Such a model 
does not have proton decays, and does not suffer from the 
doublet-triplet splitting problem.  (4) It should be noted that only 
is $G_2$ SM-like, can the breaking $G_1\times G_2\rightarrow $ SM at 
$(10-100)$ TeV occur.  Any breaking of 
$SU(5)\times SU(5)\rightarrow SU(5)$ \cite{dns} would have occurred 
above $3\times 10^{16}$ GeV.  (5) Some of the matter contents of $G_2$, 
such as the third generation can be moved into $G_1$.  Due to GMSB, the 
superpartners in this sector are very heavy $\sim 10-100$ TeV.  They 
decouple at $(1-10)$ TeV energy scale.  At this low energy scale the 
fermions, on the other hand, can form condensates due to the strong 
gauge interactions.  Dynamical fermion masses might be generated 
\cite{susytopcolor}.  In order to keep the strong GUT, it is possible 
to either introduce one more ${\bf 5}+{\bf \bar{5}}$ multiplet of $G_2$, 
which may play a role of SUSY breaking messengers \cite{kl}, or lower 
the SUSY breaking and messenger scales to be around $10$ TeV.  These 
possibilities should be studied further and are beyond the scope of 
this work.  (6) If the $SU(3)_1$ interaction is switched off, the model 
is a kind of top-flavor models \cite{topflavor}.  

This model has interesting phenomenology.  Besides the new gauge bosons, 
gauginos and Higgs particles with masses around $(10-100)$ TeV, the SM 
gaugino masses are predicted to be as heavy as $\sim 100~{\rm GeV}-1$ 
TeV.  Let us analyze the gaugino spectrum in more detail.  The full 
gaugino masses have two origins: SUSY breaking (soft masses) and 
spontaneous gauge symmetry breaking.  It has been obtained in Ref. 
\cite{kl} that the relevant mass matrix in the basis of $\lambda_r$, 
$\lambda'_r$ and the higgsino $(\psi_1-\psi_2)/\sqrt{2}$ is 
\beq
\label{12}
M_r = \left(
\begin{array}{ccc}
M_{\lambda_r}  & 0               & \sqrt{2}g_rv  \\
0              & M_{\lambda'_r}  & \sqrt{2}g'_rv \\
\sqrt{2}g_rv   & \sqrt{2}g'_rv   & 0                      \\
\end{array}\right)\,, 
\eeq
where $\psi_1$ and $\psi_2$ stand for the fermion components of 
$\Phi_1$ and $\Phi_2$, respectively.  Numerically at the scale 
$v\sim (10-100)$ TeV, $g'_r\sim 0.1$, $g_r\sim 1$.  The mass matrix 
determines two heavy states with masses 
$\sim M_{\lambda_r}\sim g_rv\sim (10-100)$ TeV, and one lighter state 
$\sim (g'_rv)^2/M_{\lambda_r}\sim 100 ~{\rm GeV}-1$ TeV.  This lighter 
state is a mixture of the $G_2$ gaugino with the higgsino.  It is 
regarded as the MSSM gaugino in this model.  On the other hand, the 
soft masses of the three generation matters are about $100$ GeV.  
Therefore in this model the SM gauginos are generally heavier than the 
sleptons and squarks.  Such a mass pattern can be tested in future 
colliders.

\vspace{1.5cm}

\acknowledgments
The author acknowledges support from the National Natural Science 
Foundation of China.

\end{document}